*Research Article*

# New Business Model for Sustainable Retail Company Using Design Thinking Concept

**[1]Anton Kurniawan, [2]Yos Sunitiyoso**
*[1,2]School of Business Management, Institut Teknologi Bandung, Bandung, Indonesia.*

*Abstract: The waste problem is still becoming a big concern in Indonesia. Waste, especially plastic waste comes from single-use packaging of daily necessities such as personal care and home care. PT. Siklus Refil Indonesia or Siklus, a retail company, comes to offer a sustainable solution of buying daily necessities by refill method. Since April 2020, Siklus has operated in the Greater Jakarta area and already impacted 20,000 customers. However, Siklus must change its new business model due to regulation from the Food and Drug Supervisory Agency (BPOM) that warned the company not to sell personal care who come in direct contact with skin. The warning impacted the decreasing customers, sales, and profit of Siklus. This research has the purpose of determining the new business model of Siklus using the design thinking concept. By this concept, this research empathizes with customers, defines customer needs, and ideates a business model. This research continues to decide the new business model by creating a matrix of stepwise selection. Then this research does a prototype business model and tests the new business model. After doing the process, Return from Home is selected as the new business model for Siklus.*

## I. INTRODUCTION

Indonesia is an emerging country in Asia that has 1.32 trillion USD in Gross Domestic Product in 2022 (World Bank, 2023). Indonesia noted a population growth rate of 1.17%, from 270 million to 273.16 million (Badan Pusat Statistik, 2023). However, during that year, Indonesia also created 19.5 million tonnes of waste and 60% ended up in landfills without any treatment (Sistem Informasi Pengelolaan Sampah Nasional, 2023). About 18% of waste produced in Indonesia comes from plastic and it is related to single-use packaging of daily necessities.

An economic system that is restoring or regenerating by purpose and design is known as a circular economy (MacArthur, 2013). A circular economy seeks to maximize resource utilization while collaboratively taking long-term social, environmental, and economic sustainability into account (Marjamaa, 2022). A circular economy must be used by the plastics industry in order to lower annual material value losses (World Economic Forum, 2016).

Jane Von Rabenau and Laksamana Sakti launched PT Siklus Refil Indonesia, often known as Siklus, on April 20, 2020. They team together to establish Siklus, a sustainable retail company that concentrates on green ways to lessen the usage of single-use plastic. The business is dedicated to giving users alternatives to single-use plastic. For domestic supplies like shampoo, soap, cooking oil, detergent, etc., the company offers a replenishment system. After three years, the areas of Jakarta, Depok, Tangerang, and Bekasi are now under Siklus delivery coverage.

During 2020-2023, Siklus will get an average revenue of IDR 500 million per month. They can support its operating expenses combined with a grant from investors total of IDR 36 billion to build refill operational and Siklus application. Siklus also give education to hundreds of moms all over Jakarta about waste issues and introduce refill solutions. Siklus also has the opportunity to open a refill pilot in Lombok, West Nusa Tenggara and Labuan Bajo, East Nusa Tenggara.

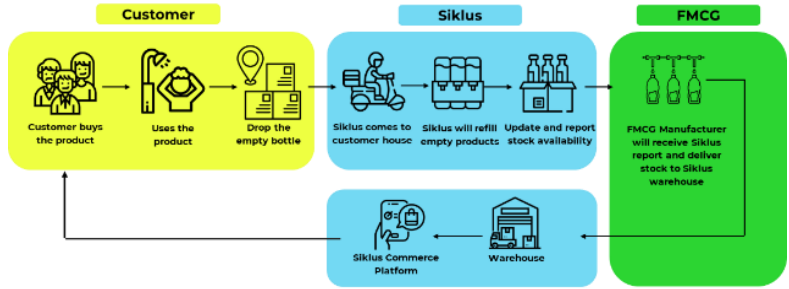

**Figure 1: Business Model of Siklus**

On April 1, 2022, Siklus got a directive from the Food and Drug Supervisory Agency (BPOM) to stop selling products, including shampoo, bath soap, and hand washing soap that comes in refills and comes into close contact with the skin. The BPOM regulation number 2 of 2020 offers socialization regarding the rules governing the monitoring of the manufacture and distribution of cosmetics. On March 23, 2023, this regulation was officially approved as BPOM regulation 12 of 2023. It is being analyzed that this regulation should be implemented on cosmetics producers such as local skin care producers, but Siklus has also been affected by regulation.

This regulation already reduces Siklus revenue by 58% due to the limitation of products that can be offered to customers. From 6.2 billion IDR in 2021 to 5.9 billion IDR in 2022 and 2.5 billion IDR in 2023, the decline in three consecutive years reflects the implication of the regulation.

Siklus then suffered profit reduction until now and created an impact on company life. The proper solution is needed to handle the situation, and a new business model is proposed while still considering the circular economy and Siklus vision as a sustainable retail company.

## II. LITERATURE REVIEW

### A) Reuse Model

One of the main components of a plastic circular economy is reusable packaging. Apart from its ability to mitigate plastic waste, it also holds promise for benefiting businesses via enhanced user experiences, cost savings, and heightened brand loyalty. The transition from single-use to reusable packaging is one of the main tenets of a circular economy for plastics since it can cut waste and plastic pollution while preserving high-value package usage.

There are four models in the reuse models that were adopted from the Ellen MacArthur Foundation.
1. Refill at home: Users refill their reusable container at home (e.g., with refills delivered through a subscription service)
2. Refill on the go: Users refill their reusable container away from home (e.g., at an in-store dispensing system)
3. Return from home: Packaging is picked up from home by a pickup service (e.g., by a logistics company)
4. Return on the go: Users return the packaging at a store or drop-off point (e.g., in a deposit return machine or mailbox)

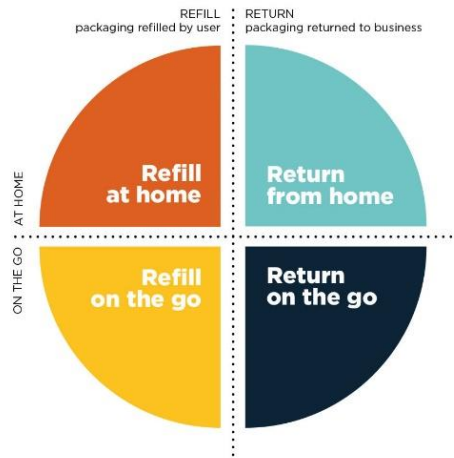

**Figure 2: Reuse Model**

*B) Design Thinking Concept*

A discipline known as "design thinking" balances what people want with what is technically feasible and what a viable business plan can convert into value for customers and potential for the market. It accomplishes this by applying a designer's taste and process (Brown, 2008).

There are five stages in the design thinking concept:

1. Empathize: The development of products or services starts with understanding the customers first.
2. Define: A deeper comprehension of the client—their ideas, emotions, experiences, and requirements—can yield more attributes (Luchs, 2015).
3. Ideate: The creators used brainstorming to translate consumer information into buildable concepts during the ideation process. The product needs to be innovative in three ways: it needs to be a profitable, viable, and desirable solution (Orton, 2018).
4. Prototype: Develop a prototype for the product that will be offered to the target customers.
5. Test: Conducted with four steps, which include test preparation, testing the system in a real-world setting with possible users, this one can be using face-to-face interviews, then documenting results in a feedback-capture grid framework, and inferring learning from the testing process (Lewrick, 2018)

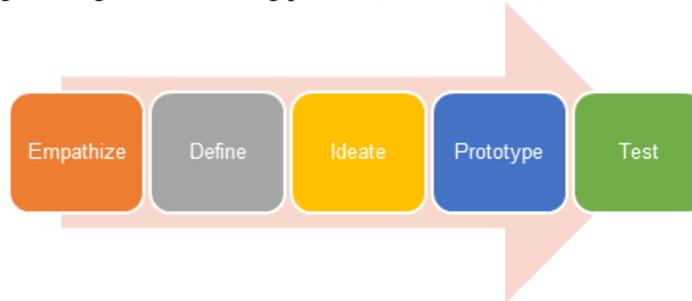

**Figure 3: Design Thinking Concept**

*C) Triple Layered Business Model Canvas*

A tool for looking into creative business models with an emphasis on sustainability is the Triple Layered Business Model Canvas (TLBMC). It expands the original business model canvas by two levels: a lifecycle perspective gives an environmental layer, and a stakeholder perspective gives a social layer (Joyce, 2016).

When combined, the business model's three layers clarify how an organization creates various forms of value, including social, environmental, and economic value. A more thorough and complete understanding of the model is created and shared with the use of this canvas tool for visualizing business models, and this, in turn, inspires creative thinking towards environmentally friendly organization models.

The TLBMC add more perspective to the economically oriented business model canvas from Osterwalder and Pigneur (2010).

This research will answer the research question of what is the new business model for tackling the low revenue of Siklus and how Siklus should implement the strategy to improve its sales.

*D) Methods*

Both quantitative and qualitative data collection techniques are used in this study. While the quantitative technique of data collecting involved distributing a questionnaire in the market survey to 100 respondents, the qualitative way involved conducting interviews with three top management of Siklus and nine Siklus customers.

For the questionnaire, here are the criteria for respondents:

1. Siklus Refill user
2. Age 18 – 72 years
3. Have made transactions in Siklus Refill

After distributing a questionnaire to 100 respondents, the customer will be classified into three personas. Then, the selected customer representing each persona will be interviewed to become a sample of the design thinking concept.

For an interview, here is the list of respondents:

**Table 1: List of Interviewee**

| No. | Name | Job Role or Customer Persona |
|-----|------|------------------------------|
| P1 | Laksamana Muhammad Sakti | Chief Operating Officer (COO) of Siklus |
| P2 | AJ Lee | Chief Executive Officer (CEO) of Siklus |
| P3 | Vinsensius Gonaldson | Cluster 1, Single Male |
| P4 | Raden Ladityarsa Ilyankusuma | Cluster 1, Single Male |
| P5 | Jan Michael Edward Tinsay | Cluster 1, Single Male |
| P6 | Andi Fachrah Savitri | Cluster 2, Young Mom |
| P7 | Agistya Maharani Joner | Cluster 2, Young Mom |
| P8 | Rani Yasmina Rosa | Cluster 2, Young Mom |
| P9 | Elisa Syawalia | Cluster 3, Single Female |
| P10 | Aubrey Maureena | Cluster 3, Single Female |
| P11 | Eriva Ramadhanty | Cluster 3, Single Female |

Then design thinking process is conducted by empathizing with customers, defining customer needs, ideating the business model, prototyping the business model, and testing the new business model. The next step is determining the new business model by creating a matrix of stepwise selection using the speed of dissemination and implementation.

The last step is creating Triple Layered Business Model Canvas (TLBMC) to summarize how Siklus should implement the strategy to improve its sales.

### III. RESULTS AND DISCUSSION

*A) Empathize*

From 100 respondents, the customer of Siklus is segmented into three personas as follows:

**Table 2: Customer Persona of Siklus**

| Persona | Single Male Ideal Iqbal | Young Mom Natural Nikita | Single Female Active Amanda |
|---------|-------------------------|--------------------------|------------------------------|
| Cluster | 1 | 2 | 3 |
| No. of Member | 27 | 36 | 37 |
| Gender | Male | Female | Female |
| Age | 18-25 y.o. | 26-35 y.o. | 18-25 y.o. |
| Education Background | S1 | S1 | S1 |
| Marital Status | Single | Married (With kids or no kids) | Single |
| Monthly Income | Rp5.5-7.5 million | Rp5.5-7.5 million | Rp5.5-7.5 million |
| Monthly Expenses | Rp3.4-5.5 million | Rp5.5-7.5 million | Rp3.4-5.5 million |
| Expenses for Personal and Home Care | Rp150-300 thousand | Rp150-300 thousand | Rp150-300 thousand |
| Time to Buy Personal and Home Care | 1x/ month | 1x/ month | 1x/ month |
| Subscription Time | 1-6 month | 12-18 month | 1-6 month |
| Last Time Buy at Sikus | 1-2 months ago | 1-2 months ago | 1-3 weeks ago |

*B) Define*

After the interview, this research can define jobs to be done, pains, and gains.

**Table 3: Customer Siklus Define Step**

| Persona | Ideal Iqbal | Natural Nikita | Active Amanda |
|---------|-------------|----------------|----------------|
| When | I want to buy home care | My personal care is empty | I want to buy personal and home care |
| I want to | Not create plastic waste | Not going outside to buy personal care | Get an affordable price and not create waste |
| So I can | Save environment | Save time | Save money |

**Table 4: Customer Jobs, Pains, and Gains**

| No | Customer Job | Customer Pains | Customer Gains |
|---|---|---|---|
| 1 | Buy personal and home care | Out-of-stock products | Products always available |
| 2 | Find the products | Cannot find the item | Have a good variance of products |
| 3 | Find in the market or Siklus | Over budget/limited budget | Affordable price |
| 4 | Find certain products that meet budgets | Still, create single-use packaging waste | No create single-use packaging waste |
| 5 | Check the product stock | Still have manual refill method | Automatic refill method |
| 6 | Check the delivery cost (if buying in Siklus) | Must wait for the motorist | Not wait for the motorist |
| 7 | Wait for the products to arrive (if buy in Sikus) | Delivery fee Rp10.000 | Less delivery fee |
| 8 | Pay for the products | D+1 delivery time | Same-day delivery time |
| 9 | Use the products | Worry about contaminating in refill method | No contamination in the refill method |
| 10 | Dispose of the packaging waste | Less discount and promo | More discounts and promo |

*C) Ideate*

This research uses Substitute, Combine, Adapt, Modify, and Put to Other Uses, Eliminate, and Rearrange (SCAMPER). In this step, to determine the new business model of Siklus, the Adapt and Modify perspective will be explained further:

1. Adapt (What other concepts does it raise? Exists anything comparable exist that could be used to solve the current issue? Have comparable circumstances occurred before?)
2. Modify (What change might be made? Is it possible to alter the meaning? In what way could the business model be modified? What can be made more of? What can be cut down? What has to be updated? Is it possible to make it larger? Is it small enough?)

Then, three alternatives of the new business model (A: Refill on the Go; B: Return from Home; C: Return on the Go) are scored during several parameters of ability to solve pain and gain customers:

**Table 5: Pains-Gains-Solution**

| No | Pains | Gains | A | B | C |
|---|---|---|---|---|---|
| 1 | Out-of-stock products | Products always available | 0 | 1 | 0 |
| 2 | Cannot find the item | Have a good variance of products | 0 | 1 | 0 |
| 3 | Over budget/ limited budget | Affordable price | 1 | 1 | 1 |
| 4 | Still, create single-use packaging waste | No create single-use packaging waste | 1 | 1 | 1 |
| 5 | Still have manual refill method | Automatic refill method | 0 | 1 | 1 |
| 6 | Must wait for the motorist | Do not wait for motorist | 1 | 0 | 1 |
| 7 | Delivery fee Rp10.000 | Less delivery fee | 1 | 0 | 1 |
| 8 | D+1 delivery time | Same-day delivery time | 1 | 0 | 1 |
| 9 | Worry about contaminating in refill method | No contamination in the refill method | 0 | 1 | 1 |
| 10 | Less discount and promo | More discounts and promo | 1 | 1 | 0 |
| | | Total | 6 of 10 | 7 of 10 | 7 of 10 |

Then, the step continues with determining which business model goes through a prototype test (Lewrick, 2018). By creating a matrix of stepwise selection using the speed of adaption and dissemination matrix in the first step and then the financial feasibility and implementation matrix in the second step, the business model is evaluated.

**Table 6: Speed of Adaption and Dissemination Matrix**

| | | Speed of Dissemination | |
|---|---|---|---|
| | | Quick | Slow |
| Speed of Adaption | Quick | Develop the idea further | Continue to develop if we can rollout |
| | Slow | Continue to develop only if we can influence the buy-in of decision-makers who influence the rollout. | Find alternative ideas |

**Table 7: Financial Feasibility and Implementation Matrix**

| | | Implementation | |
|---|---|---|---|
| | | Yes | No |
| Financial Feasibility | Yes | Build a prototype yourself | Search for partners for implementation |
| | No | Revise/ simplify the idea | Discard the idea |

For option A: refill on the go, the response from customers interested in this business model is 3.57 out of 5, when 5 stands for strongly interested. In step 1, this business model can create slow adaption and slow speed of dissemination due to the obligation to partner with supermarkets and warung. The idea is determined not to continue into the prototype step.

For option B: return from home, the response from customers interested in this business model is 4.00 out of 5, when 5 stands for strongly interested. In step 1, this business model can be quick to adapt and quick to disseminate due to the facility and resources acquired by Siklus just to shift a little bit to accommodate the new business model. In step 2, the financial is feasible due to the higher margin from a supplier and the cleaning fee also as a revenue stream. The implementation is also feasible, so the idea is processed to the prototype step.

For option C: return on the go, the response from customers interested in this business model is 3.21 out of 5, when 5 stands for strongly interested. In step 1, this business model can create slow adaption and slow speed of dissemination due to the obligation to partner with supermarkets and warung. The idea is determined not to continue into the prototype step. However, it can be a further step of Siklus's business model if, in the future want to expand to a larger scale.

The new business model of return from home is selected. Siklus will propose sustainable packaging and products as a value proposition. By providing bottle collecting and cleaning, all the activities conducted by Siklus can be accounted for.
Siklus will create partnerships with packaging producers, do bottle collecting and cleaning, and retail selling. Siklus major activities are funding, training new employees, and creating a warehouse. Siklus also need to resource product manufacture while opening partnerships with wholesalers, packaging suppliers, and the government.

The scheme of the new business model, return from home is as follows:

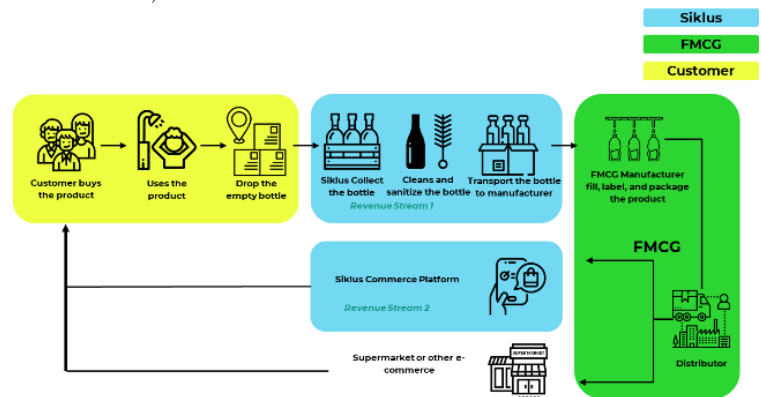

**Figure 4: New Business Model of Siklus**

*D) Prototype*
This research continues to create prototypes of a return on home new business model by providing aluminium bottles for Safi shampoo. In first buying, the customer will buy the bottle and the product. In the next buying, the customer only buys the product. The old packaging will be taken by Siklus, cleaned, and delivered to the supplier. It solves two problems: from the customer, they don't create waste from packaging and don't feel difficult to treat the waste; from the FMCG supplier, in this case Unza Wipro, it reduces the cost of packaging and can tap into Siklus customers. It is become the first pilot project to gather feedback from customers.

In this prototype, Siklus customers can buy Sampo Safi Extracare 500 ml and an aluminium bottle with Rp72.000/pcs. For the next purchase, Siklus customers can get a new product with Rp50.400/pcs but with the requirement to return the packaging. Siklus customer can order via application and marketplace Siklus, wait at their house, and accept the product that will be delivered by Siklus. For reference, in the marketplace, Sampo Safi Extracare 160 ml with regular packaging is Rp22.100/pcs.

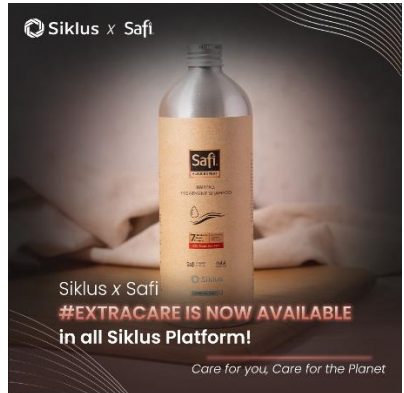

**Figure 5: Siklus x Safi Pilot Project**

Here is the prototype result:

**Table 8: Prototype Result**

| Market segment | Value proposition |
|---|---|
| Urban millennial | No single-use packaging waste |
| Living in Jabodetabek | Affordable price |
| | No manual refill method |
| Revenue model | Growth model |
| Purchase fee | Collaborate with brand |
| Cleaning fee | Scalable to other cities outside |
| Plastic credit | Jakarta |

*E) Test*

In the testing stage, held in January 2024, Safi shampoo is obtained and can be purchased by customers. Then, this research use a feedback grid to gather feedback from 100 customers in a structured grid to improve the new business model.

The average score of feedback from 100 customers is 3.48 out of 5, when 5 stands for strongly interesting. The feedback from customers is also collected and divided into four areas: likes, wishes, questions, and ideas. The feedback from customers is developed to make the new business model more sustainable.

Here is the summary of the test result:

**Table 9: Test Result**

| Likes | Wishes |
|---|---|
| • Interesting packaging: there are not many products that use aluminium bottle packaging<br>• Siklus helps to reduce bottle waste at home, and there is some pride when doing good for the environment.<br>• Reducing the price in the next purchase is quite effective in increasing customer loyalty.<br>• Sustainable packaging that can be used more than once<br>• Fast response, informative motorist, Siklus, keep up! | • Siklus can expand outside Jakarta<br>• Have many more variance of products<br>• Delivery service 1-2 hours after orders<br>• Add refill station and increase delivery area<br>• Have smaller size of products |
| Questions | Ideas |
| • How is the process after the packaging is returned back? Is there more sustainable? | • Siklus conducted a discussion session on the new model business with the customer. |
| • If I have warung, can we get a partnership to refill the station? <br> • Why did the pilot project start with shampoo? Some environmentalist now promotes no-shampoo activism. <br> • If Siklus applies the system, will there be a sanction if the customer doesn't return the packaging? <br> • How does Siklus guarantee the origin of the products, especially personal care? | • Siklus can create an offline store with a refill station at Jakarta Mall <br> • Vending machine to sell products and return reused packaging <br> • Socialization with RT/RW to promote a sustainable lifestyle, not only seeking the cheapest product <br> • Every reuse gets points, creates a community Whatsapp, and keeps growing |

For the new business model of Return From Home, this research creates Triple Layered Business Model Canvas as follows:

1. Economic Business Model Canvas

Siklus will utilize its customers from Business to Business (B2B) and Business to Customer (B2C) via the Siklus application, e-commerce, and social media.

From a financial perspective, the cost of bottle collection and cleaning can be covered by the partnership with B2B and selling the product retail in B2C so, making it a sustainable and profitable business.

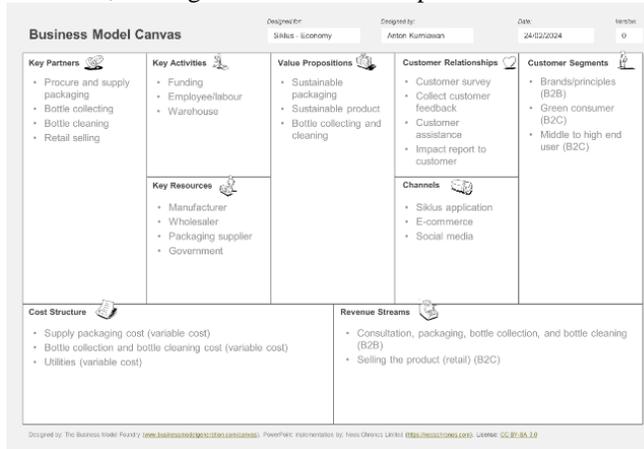

**Figure 6: Economic Business Model Canvas**

2. Environmental Business Model Canvas

Siklus can save 0.7 kg of plastic waste per day from every customer. Align with that, the Environmental, Social, and Governance (ESG) report can be published by Siklus every year to show the impact created by Siklus.

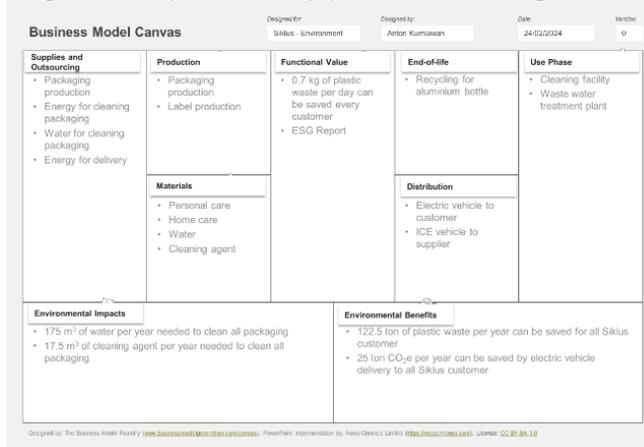

**Figure 7: Economic Business Model Canvas**

3. Social Business Model Canvas

   Siklus will commit to the vision of making everyday necessities more affordable. The cost of packaging will be reduced from the product and make the price lower than before.

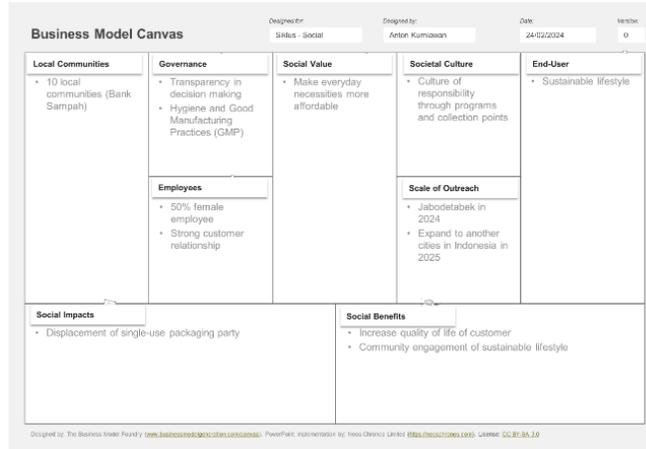

**Figure 8: Social Business Model Canvas**

## IV. CONCLUSION

By using the design thinking concept, the new business model of Siklus is to return from home. For the new business model, a triple layer of business modal canvas has already been created.

A number of recommendations could be given as management implications to monitor the project dynamics and further exercise the business solution. Siklus Return from Home must be implemented as soon as possible with several brands to get real market data and customer purchase behaviour. Siklus should maintain a good relationship between investor and supplier to see another opportunity in the business model of Siklus Return on The Go.

Further research can be conducted to educate and change the Indonesian mindset, especially low-middle income society, to use less single-use packaging and move into a more sustainable way of purchasing daily necessities.

## V. REFERENCES

[1]  Agenda, I. (2016, January). The new plastics economy rethinking the future of plastics. In The World Economic Forum: Geneva, Switzerland (Vol. 36).
[2]  Badan Pusat Statistik (2023, September 7). Laju Pertumbuhan Penduduk. https://www.bps.go.id/indicator/12/1976/1/laju-pertumbuhan-penduduk.html
[3]  Cooper, R., Junginger, S., & Lockwood, T. (2009). Design thinking and design management: A research and practice perspective. Design Management Review, 20(2), 46-55.
[4]  Eberle, B. (1996). Scamper on: Games for imagination development. Prufrock Press Inc..
[5]  Ferreira, B., Silva, W., Oliveira, E., & Conte, T. (2015, July). Designing Personas with Empathy Map. In SEKE (Vol. 152).
[6]  Food and Drug Supervisory Agency (BPOM) Regulation no 2 of 2020 concerning the Supervision of Production and Distribution of Cosmetics
[7]  Food and Drug Supervisory Agency (BPOM) Regulation no 12 of 2023 concerning Supervision of the Manufacture and Distribution of Cosmetics
[8]  Jones, P., Hillier, D., Comfort, D., & Eastwood, I. (2005). Sustainable retailing and consumerism. Management Research News, 28(1), 34-44.
[9]  Lewrick, M., Link, P., & Leifer, L. (2018). The design thinking playbook: Mindful digital transformation of teams, products, services, businesses and ecosystems. John Wiley & Sons.
[10] Lockwood, T. (2010). Design thinking: Integrating innovation, customer experience, and brand value. Simon and Schuster.
[11] Luchs, M. G. (2015). A brief introduction to design thinking. Design thinking: New product development essentials from the PDMA, 1-12.
[12] MacArthur, E. (2013). Towards the Circular Economy. Journal of Industrial Ecology, 2(1), 23-44.
[13] MacArthur, E. (2019). Reuse Rethinking Packaging. Ellen MacArthur Foundation.
[14] Marjamaa, M., & Mäkelä, M. (2022). Images of the future for a circular economy: The case of Finland. Futures, 141, 102985.
[15] Matthews, J., & Wrigley, C. (2017). Design and design thinking in business and management higher education. Journal of Learning Design, 10(1), 41-54.
[16] McKinsey (2023, September 7). Sustainability in Packaging. https://www.mckinsey.com/industries/paper-forest-products-and-packaging/our-insights/sustainability-in-packaging-global-regulatory-development-across-30-countries
[17] Orton, Kristann. (2018). Human-Centered Innovation. Inceodia.

[18] Osterwalder, A., Pigneur, Y., Oliveira, M. A. Y., & Ferreira, J. J. P. (2011). Business Model Generation: A handbook for visionaries, game changers and challengers. African journal of business management, 5(7), 22-30.

[19] Patton, M.Q. (1999). Enhancing the quality and credibility of qualitative analysis. Health Sciences Research, 34, 1189–1208.

[20] Saunders, M. N., & Townsend, K. (2016). Reporting and justifying the number of interview participants in organization and workplace research. British Journal of Management, 27(4), 836-852.

[21] Sileyew, K. J. (2019). Research design and methodology. Cyberspace, 1-12.

[22] Sistem Informasi Pengelolaan Sampah Nasional (2023, September 7). Capaian Kinerja Pengelolaan Sampah. https://sipsn.menlhk.go.id/sipsn/

[23] Slovin, E. (1960). Slovin's formula for sampling technique. Retrieved on February, 13, 2013.

[24] Statista (2024, February 17). Beauty and Personal Care Market in Indonesia. https://www.statista.com/outlook/cmo/beauty-personal-care/indonesia

[25] Statista (2024, February 17). Home and Laundry Care Market in Indonesia. https://www.statista.com/outlook/cmo/home-laundry-care/indonesia

[26] Svoboda, S. (1999). Note on life cycle analysis. Environmental management: Readings and cases, 217-227.

[27] Wirtz, B., & Daiser, P. (2017). Business model innovation: An integrative conceptual framework. Journal of Business Models, 5(1).

[28] Worldbank (2023, September 7). GDP Current US$ Indonesia. https://data.worldbank.org/indicator/NY.GDP.MKTP.CD?locations=ID